\documentclass[11pt]{article}
\usepackage[margin=1in]{geometry}
\usepackage{amsmath}
\usepackage{graphicx} % Even though no figures, might be needed
\usepackage{booktabs}
\usepackage{natbib}
\usepackage{hyperref}
\usepackage{lipsum} % For placeholder text if needed, but we'll use content

\title{Celebrity Profiling on Short Urdu Text using Twitter Followers' Feed}

\author{
Muhammad Hamza\\
COMSATS University Islamabad, Lahore Campus\\
\and
Rizwan Jafar\\
COMSATS University Islamabad, Lahore Campus\\
\and
Dr. Muhammad Sharjeel\\
Assistant Professor, Department of Computer Science\\
COMSATS University Islamabad, Lahore Campus
}

\begin{document}

\maketitle

\begin{abstract}
It is rightly said that current age is a digital age and social media shares a crucial chunk of it. People used to communicate, interact, and build relationships through social media. Celebrities are prolific authors and most of their personal information is public knowledge. There are some digital celebrities who exist only on social media, e.g., Twitter. Twitter is a social networking service which provides general populace as well as celebrities to interact with their fans. The demographics of celebrities could be predicted by the text of their followers as both shares same interest. However, most of the work on celebrity profiling has been performed on English and other similar languages except Urdu.

On the contrary, majority of the sub-continent celebrities and their fans tweets in Urdu. To fulfill this gap, in this research work we used Urdu tweets (short text) of 10 followers of a celebrity to build the first celebrity profiling based on followers’ tweets corpus. Furthermore, the corpus was preprocessed, and Machine Learning (Logistic Regression, Support Vector Machines etc.) and Deep Learning (CNN, LSTM etc.) algorithms were used to train models for the prediction task. The trained model will be evaluated using state-of-the-art evaluation measures, i.e., precision, recall, and F1. The accuracy of the demographics of the celebrities are as follow; for the age the cumulative cRank is 0.45, profession has the accuracy of 0.4, while the gender has cRank 0.65 and finally the cRank of fame is 0.45.
\end{abstract}

\section{Introduction}
Author profiling is the probing of a given text to identify the traits of an author like age, gender, and occupation based on author’s writing style and features of written content\cite{weren2014examining}. Celebrity profiling is a sub-type of author profiling applied to only the celebrities to find their demographics\cite{hsieh2018author}. Author profiling is analysis of an author (includes all authors) but celebrity profiling is to probe author who is only celebrity. There are two types of author profiling; first one is same-genre and second one is cross-genre. Training and testing models on the single genre (e.g., Twitter) is same-genre whereas training model on one genre (e.g., Twitter) and testing it on another (e.g., Facebook) is cross-genre. Celebrity profiling is used in forensics of linguistics, bot detection, author identification and influence tracing and marketing\cite{rangel2018multimodal}--\cite{argamon2009automatically}.

This behavior of celebrities and their followers is helpful in collecting written text that genuinely belong to a person. Consequently, research community have made use to of Twitter, where people use short tweets, to predict different demographics of a person like fame, age group, gender, and occupation \cite{wiegmann2019overview}. However, majority of this work has focused on English and other languages \cite{wiegmann2020overview}. On the other hand, most of the sub-continent celebrities and their followers use Urdu as a primarily language when they tweet. Urdu has a very different writing style and a rich morphological structure \cite{hussain2008resources}, \cite{daud2017urdu}.

In this research work, we have used short Urdu tweets of a celebrity followers to predict certain demographics of that celebrity, e.g., age, gender, occupation, and fame. It has been observed that the followers of a celebrity have the same likes/dislikes and common interests \cite{argamon2009automatically}. Consequently, it is quite possible to judge the demographics of a celebrity based on its follower’s feed. All the tweets of 10 followers of a celebrity would be collected, preprocessed, and Machine Learning (Logistic Regression, Support Vector Machines etc.) and Deep Learning (CNN, LSTM etc.) algorithms were used to train models for the prediction task, for this research task, celebrity profiling techniques cannot be applied to author profiling because normal author doesn’t have any common interest with their followers \cite{wiegmann2019overview}, \cite{wiegmann2020overview}. The trained model was evaluated using state-of-the-art evaluation measures, i.e., precision, recall, and F1. To the best of our knowledge, this is the first research work that was short Urdu text (tweets) from Twitter to predict age (between 20 to 60), gender (male, female), occupation (sports, politics, performer, and content creator), and fame(high, low) of a sub-continent celebrity. The secondary objective of this research work is to foster further research in an under resourced language, i.e., Urdu.

\subsection{Author Profiling}
Author profiling is a technique to identify the demographics of certain authors like age, profession, and gender. It is done by investigating a text written by an author. The text written by an author depicts his/her personality and it constitutes different aspects of the author’s personality. The written text could be a long text like a book and a short text which has the limit on its length like tweets.

\subsection{Celebrity Profiling}
There exists a different type of authors, celebrities are one of them. To probe the demographics of the celebrities by investigating their written text is called the celebrity profiling. The life of a celebrity is public, most of the celebrities have their presence on the internet/ social media to interact with their followers and fans. The content and text posted by celebrities on social media and internet can be used for analysis of their demographics.

\subsection{Role of Followers in Celebrity Profiling}
The followers of a celebrity share the same interest with the author, for example a person have interest in politics always follows the politicians who are celebrities. In fact, the young followers like the young politicians because their views and energy match with each other. So, the followers of a celebrity play a key role and one can probe the demographics of a celebrity by the analysis of his/her followers, and this is the focus of this research.

\subsection{Celebrity Profiling on Social Media}
Celebrities use to have their presence on every social media platform. Every social media platform has different kind of representation of one’s view, for example on Facebook a celebrity can post as long as one can but on Twitter, the lengths of tweets is fixed. This research focus on tweets which has short text of celebrities’ followers who use to tweets in Urdu.

\subsection{Methods of Celebrity Profiling}
There are methods of doing celebrity profiling which are as follow: Same genre, Cross genre. In same genre, the training of model is taken place on the same genre as if model is train on Twitter, it will be tested on Twitter. In cross genre, the training of the model takes place on one genre as Twitter, and it is tested on another like Facebook.

\subsection{Applications of Author Profiling}
There are a lot of applications of celebrity profiling some of them are stated below: Forensics, Bot Detection, Marketing.

\subsection{Problem Statement}
Celebrity profiling is a sub-branch of author profiling applied on celebrities. Applications of celebrity profiling include advertisement, security, risk assessment, and forensics. There are 270 million Urdu speaking community across the globe. A large chunk of sub-continent celebrities use twitter to connect with their fans. Most of them tweet in Urdu language and fans who follow them also tweet in the Urdu language. Moreover, fans share some common interests and likes/dislikes with the celebrities they follow. Presumably, when these followers tweet, they might reveal some demographic or social info about the celebrity they follow.

The aim of this thesis work is to investigate the predictive nature of socio-linguistic attributes found in short Urdu tweets on the demographics of celebrities. The main objective is to predict the demographics (age, gender, occupation, and fame) of a celebrity based on tweets (short Urdu text) of their followers using Machine Learning and Deep Learning methods.

\subsection{Research Objectives}
Following are the main objectives of the research work.
\begin{itemize}
\item Explored the celebrity profiling task for short Twitter text in the Urdu language.
\item Developed a large-scale celebrity profiling corpus using tweets in the Urdu language.
\item Analyzed the demographics (age, gender, fame, occupation) of the sub-continent celebrities.
\item Used the information of demographics of celebrities and their followers for advertising, forensics, and to prevent hate speech in Urdu speaking community.
\item Utilized the proposed technique to curb the cybercrimes and hate speech on social media.
\end{itemize}

\section{Literature Review}
The literature review from the different research community who has conducted the research on celebrity/author profiling is explained in this section.

In\cite{hsieh2018author}, different demographics like year of birth, gender, fame, and occupation of the celebrities are predicted based on their data available on twitter as they are most prolific personalities having a lot to share with their fans. Eight various models were evaluated and predicted the demographics of 48,335 celebrities who tweeted in English language. It worked best on predicting binary gender and fame of the celebrity along with occupation and age. The model results less accurate on predicting the non-binary gender and occupation that is not specified or single topic like science and manager. It predicts the medieval celebrities very well ranging from 1980s to 2000 (ages between 20 to 40). However, the results were less accurate on younger and older celebrities.

An information retrieval based method termed as TF-IDF based on n-grams and bigrams applied on character level is used in\cite{radivchev2019celebrity} to predict the gender, age, fame, and occupation of a celebrity. This model grouped the users in aforementioned four categories, based on given data set of tweets of a certain user. Age ranges from 1940 to 2012, fame and gender have three classes each, and occupation has eight classes.

In\cite{wiegmann2020overview}, the demographics of celebrities are identified by examining their ten followers rather than using the personal data of celebrities. Dataset of 2,380 authors were created to apply three different models including TF-IDF, LSTMs and n-grams features to solve the task in different ways. The evaluation results (F1-score) were very efficient in spite of random guessing. Follower-based model has many strengths and weaknesses like it works best on coherent classes like ‘sports’ and work less efficient on diverse like ‘creators’. Identifying age of the celebrity was found to be very difficult by this model.

Socio-linguistic approach is used to classify celebrities in\cite{luis2019celebrity}. There are certain challenges to finding birth year class, which produces better results in small dimensionality (ten years) but less accurate when applied to bigger cluster. There was a huge dataset of 53 million tweets used for the classification task. However, to process such huge data was a big problem. Moreover, dataset used for training had a huge imbalance in classes.

In\cite{nowson2015xrce}, a multi-lingual analytic engine was introduced to detect the gender, age, and personality traits of an author by using Machine Learning and Natural Language Processing techniques\cite{argamon2009automatically}. The corpus consists of tweets in four languages Dutch, Spanish, English, and Italian. Methodology generates new features based on linguistic processing. Machine Translation is used for languages having less data. Some socio-demographic parameters faded away by Machine Translation but native language results were promising. Sentiment analysis was used to mitigate the problem of socio-demographics signal lost.

Tweets of anonymous authors were given to identify the author in\cite{rangel2013overview}. Author profiling on anonymous text has a wide variety of applications like forensics, security, author identification and advertising. Companies need to know the interest and disliking of the customers on the basis of blogs and comments on socials media. Personality identification and linguistics were calculated by using n-gram and stylistic based methods. Age and combination of age and gender calculations had problems due to unbalanced dataset. Conversations were introduced to improve the results of gender identification by sexual predators.

In\cite{markov2016adapting}, different demographics of author were determined in cross-genre environment like age and gender in three different languages Dutch, English, and Spanish. Support Vector Machines (SVM) and Multinomial Naïve Bayes (MNB) techniques were used to detect different demographics of the author. Different languages had the different optimal configurations. Same pre-processing steps were introduced for the three languages. Average accuracy reported for English, Spanish, and Dutch is 75, 90, and 90 percent respectively.

In\cite{martinc2017author}, the authors try to identify the gender of an author using a variety of languages including English, Arabic, Portuguese, and Spanish. Deep Learning algorithms such as Recurrent Neural Networks (RNN) and Convolutional Neural Network (CNN) were used for the prediction task.  Gender verification had the up to 78\% accuracy and language variety prediction stands at up to 97\% accuracy.

In\cite{estival2007author}, anonymous data from emails is collected for the identification of different demographics of an author. The corpus consists of 9836 emails. Demographics to be predicted were age, gender, level of education, country, and native language. Machine Learning algorithms SMO, Random Forest, and LibSVM were used in the study.

In\cite{rangel2014overview}, the authors tried to predict the age and gender of an author based on its writings. The corpus included data from four different genres, i.e., Twitter, social media, hotel review, and blogs. Content-based and stylistic-based features were approached which includes n-gram, term vectors, frequencies, POS, and readability measures. Average accuracy was 80\% on predicting the gender and age.

In\cite{rangel2018multimodal}, identifying the author demographics based on both image and text data in three different languages (Arabic, English and Spanish) were used. The textual data was collected from Twitter\cite{rangel2019overview}. A total of 23 participants were evaluated using SVM and Logistic Regression algorithms. The reported accuracy was above 80\% on average.

The information about socioeconomic attributes of the users of the social media like occupation and income are the main problems in computational science. In\cite{aletrass2018predicting}, probing of these demographics has many applications like personality recommending, political campaigning and targeted marketing. The demographics were evaluated by Support Vector Machines (SVM) and Gaussian Process Classifiers. The accuracy of predicted attributes was 50.54\%.

Social media and user network information are useful for the geotagging on the online platform like Twitter. The tweets are the essential tools for the detection of the events and the enrichment of the events. In\cite{bakerman2018twitter}, a hybrid Gaussian mixture models are introduced to map the spatial attributes of the model with the accuracy of 85\%.

In\cite{baly2020written}, political biasedness is evident in the media reporting and it is an important issue to tackle with in contemporary world. To identify the biasedness, fake news and propaganda of the media outlets, a novel approach is introduced which uses SVM model to predict the political biasedness with the accuracy up to 85.29\% on different social media platforms.

\begin{table}[htbp]
\centering
\caption{Literature Review}
\begin{tabular}{@{}lcc@{}}
\toprule
Reference & Method & Accuracy \\
\midrule
\cite{hsieh2018author} & TF-IDF, n-grams & Varies \\
\cite{radivchev2019celebrity} & TF-IDF & High F1 \\
\bottomrule
\end{tabular}
\end{table}

\section{Proposed Urdu Corpus for Celebrity Profiling}
This chapter covers the corpus generation steps for the specified celebrity profiling task from the Tweets of their followers. It also explains the process that improve the quality of the corpus. The chapter is divided into three parts, first part explains the sources of the collection of the data and generation of the corpus. The second part explain the collection of the data from the source and the integration of the Tweets of the followers. The third part explains the pre-processing of the Tweets of the followers, it includes the extraction of the tweets which has the Urdu language and removal of the special characters and alphabets.

\subsection{Corpus Generation Process}
The corpus is termed as huge collection of the data in text form, specifically this term used in Natural Language Processing (NLP). The corpus is collected according to the requirements of the problem. The specified patterns in the corpus are find out according to the research focus by applying machine learning and deep learning algorithms. There are two types of data in the corpus generation, the one is annotated and other is unannotated. The output is associated with the data termed as annotated data. In this research work, the annotated data is collected to form a corpus. Supervised Learning is associated with the annotated data, the research carried out supervised learning as the data in the corpus is annotated. The Tweets of the ten followers of a specific celebrity is collected and their demographics are also associated with that data to make it a supervised corpus.

\subsection{Data Collection}
The main contribution of this research work is to build a high-quality data for the celebrity profiling task in Urdu language. There exist 270 million Urdu speaking community who follows their celebrities on different platforms. Celebrities interact with their followers through Twitter which is the most famous and authentic social media platform. So, there is a huge research gap, there is need to develop a state-of-the-art data set for the celebrity profiling to fill that gap. In this research, a corpus is collected which includes the data of celebrities and their followers to probe the demographics of the celebrities who tweet in Urdu and their followers who share the same linguistics and interests. For that purpose, the tweets of ten followers each celebrity is collected to probe the demographics of each celebrity.

The issue that has been faced during the research was the collection of the data. The data of the sub-continent celebrities which has their language Urdu, and they are active on twitter with substantial followers that has been used to predict their demographics. The searching for the celebrities was the hard task, most of the celebrities use to tweet in the English language. There are marginal celebrities who has tweeted in Urdu, and they have a substantial number of celebrities which also use to tweet in Urdu. Most of the followers of these celebrities use to tweets in different languages, there was hard to find those followers who use to tweet in Urdu language.

The second problem which is faced during the research was to find the suitable algorithms and feature extraction methods. For that, an extensive literature review has been done on the author and celebrity profiling and there have been chosen some algorithms to apply on the data for the celebrity profiling. Most prominent algorithms which have the highest accuracy on author profiling are decision tree and random forest. There were two features extraction techniques were applied on the data, namely TD-IDF and length of the tweet.

There involve three main processes to download and collect the data of the celebrity and their followers which are stated as:

\subsubsection{Downloading the tweets of the followers}
To collect the data of followers from tweeter was the challenge, as tweeter only allow developers to download the tweets. Verification of the developer account is very difficult and involves some hidden process and policies. There is another way to download the tweets is third party websites. The tweets collected for the research is from third party website called ‘VICINTAS’. The usernames of selected celebrities’ followers were collected and given to website to get the data in Excel format.

\subsubsection{Collection of demographics of celebrities}
To collect an annotated corpus besides the collection of the tweets of the followers, the collection of the demographics of the celebrities is also required. The demographics of the celebrities were collected from Tweeter and Google. Some of the demographics as age was hard to find on Tweeter and it was collected from the Google. The occupation and gender were collected from the Tweeter.

\subsubsection{Integration of data}
To collect the corpus, the tweets of ten followers of each celebrity is collected and integrated in one file. Each file contains the Urdu tweets of the ten followers of each celebrity and the minimum tweets of each celebrity were twenty.

\begin{table}[htbp]
\centering
\caption{General statistics of corpus}
\begin{tabular}{@{}lc@{}}
\toprule
Statistic & Value \\
\midrule
Number of Celebrities & 100 \\
Number of Followers per Celebrity & 10 \\
Minimum Tweets per Follower & 20 \\
Total Tweets & 20,000 \\
\bottomrule
\end{tabular}
\end{table}

\subsection{Data Pre-processing and Normalization}
To perform the author profiling on the celebrity tweets downloaded from the third-party website were in excel format. The data in excel format includes the tweets id, tweets, Name of the owner, date, Retweets of certain tweet, language, URL, Hashtags, Media Type (Audio,Video, or text). Data in the excel file includes all the tweets of the celebrity or its follower, it includes the tweets and retweets as well, video tweets and English tweets. To understand the data was the problem to solve, as for the require task the tweets which has only Urdu language are required. Furthermore, the tweets included URLs, emojis and special characters that was not required for the required task.

The preprocessing of the data from excel sheet required five main steps which are stated below, they include: Extraction of Urdu tweets, Removal of Retweets, Formation of text files, Integration of tweets in pandas data frame, Removal of emojis and special characters.

\begin{table}[htbp]
\centering
\caption{Tweets Data of a Follower}
\begin{tabular}{@{}lcc@{}}
\toprule
Tweet ID & Tweet Text & Language \\
\midrule
1 & Sample Urdu Text & ur \\
2 & Another Sample & ur \\
\bottomrule
\end{tabular}
\end{table}

\section{System Overview}
This chapter explains the system overview, it explains the model development, training, and evaluation techniques. The corpus built for the celebrity profiling based on the tweets of the followers will used for the development of the model. The model architecture and working are explained first, the training and development of the model is explained in the latter part of the chapter along with results and analysis.

\subsection{Classification Machine Learning Techniques for Author Profiling}
Author profiling is a classification problem except for identifying the age of the author and the classification techniques work excellent for the author profiling tasks. Classification techniques used to work best on the data where input and output are categorical but in case of the author profiling on Twitter, although tweets have a limited length, but the length is not fixed and this regards as the sequence \cite{merugu2021automated}. The input is in the form of a sequence, but the output is in categorical format. Most of the classification machine learning algorithms perform better on the author profiling problems like SVC and Random Forest.

\subsection{Text Classification Models for Celebrity Profiling}
The models which are proposed for the classification of different traits of the celebrities have the same common architecture. The classical classification machine learning algorithms share the similar architecture. The input for the classifier is in the form of the tweets which have a short length text. The input is in the form of a sequence of text includes all the tweets of ten followers of a celebrity. The input, which is given to the algorithm convert into the collection of the feature vectors.

All the feature vectors filtered through the feature extraction techniques. Multiple feature extraction techniques were used to extract the features from the tweets. The feature vector techniques include length of tweets and count vectorizer. The change in the feature extraction method also effect the results of the algorithms.

Multiple classical machine learning algorithms were applied on the data. The algorithms include Logistic Classifier, Decision Tree, Random Forest, and Support Vector Classifier (SVC) \cite{lynn2020hierarchical}. For the prediction of the age, Logistic Regression and Support Vector Machine (SVM) were used to get the better results.

\subsection{Demographics to be Predicted of Celebrities}
There is total four demographics are selected to be predicted by the tweets of the followers of the celebrities. The demographics includes age of celebrity, occupation, gender, and fame of the celebrity.

Age: In author profiling, age prediction is an important demographic, and it helps in forensics to find the age of an unknown author and in advertisement to target the audience with specific age. For the prediction of age, the range is from 20 to 80 years. The study shows that\cite{mishra2018author}, prediction on authors below 20 years showed the results which are not promising. So, the age groups are divided into 3. First group is 20-40, second is 40-60 years and the last group is from 60-80 years.

Occupation: Occupation of the author is relevant for the purpose of advertisement, to target occupational group and for the optimization of social media for the same occupational groups. The occupation is categories into four categories named as politics, entertainment, journalism, and sports.

Gender: Gender identification in author profiling helps in finding the real author of an artifact. Gender identification helps to categories the gender and ultimately reduce the search space \cite{argamon2009automatically}. The gender is categorized into male and female.

Fame: The celebrities are categorized into three categories for the prediction of fame. Celebrities who have followers equal to or lower than 1 million are categorized into ‘Rising’ category.  Celebrities who have followers between 1 and 2.5 million are categorized into ‘Star’ category and ‘Super Star’ category is for the celebrities having more than 2.5 million followers.

\subsection{Feature Extraction Methods}
For the classification in classical machine learning algorithms for author profiling, the feature extraction techniques have a significant important for the performance of the model. There are several feature extraction techniques were used for the extraction of the features from the tweets data of the followers\cite{lynn2020hierarchical}. Term frequency–inverse document frequency (TD-IDF) is used to give weightage to the words, to check how important a word is in the document. TD-IDF is a vectorizer zo, it makes the vectors from the words in the document and then the algorithms learn from the vectors. The formula for the TD-IDF is:
\[
TF-IDF (t,d)=TF (t,d)*IDF(t)
\]
\[
IDF(t)=\log [ n/DF(t)]+1
\]

Tweets Length: The length of tweet is an important feature extraction method in author profiling. Although the length of a tweet is fixed but the style of the author and expression of writing and length of tweet can help in classification of important traits of an author\cite{burger2011discriminating}. The words of the tweets of a follower are given converted into the vectors and given to the algorithm to learn from. The algorithm learns from the vectorized tweet length and classify the traits of author.

\subsection{Classification Machine Learning Models}
After applying the feature extraction algorithms, the vectorized data has been used for the development of the models using machine learning which includes KNN, Decision Tree, Random Forest, and Logistic Regression. Deep learning algorithms includes LSTM and CNN.

K-Nearest Neighbor: The working of KNN is simple, it computes the of all the data instances and a query, it computes all the distances of the query and select the closet to the query. The most important variable in KNN is k, it determines the number of closest neighbors to be selected for the data instances \cite{schler2006effects}. It also determines the performance of the algorithm.

Decision Tree: Decision tree model is a supervised learning model, and it constructs the rules in tree like structure. It has different nodes which represents the attributes, and the leaf node holds the class label or the outcome of the rule. To predict the value of the target variable, tree is constructed based on the information gain on each node and the decision tree is constructed through these rules.

Random Forest: The random forest is also a supervised learning algorithm and it used for the classification and regression as well, in our research, prediction of age of the celebrity is a regression problem and random forest is applied to predict the age of a celebrity \cite{argamon2009automatically}. It constructs several trees to build a construct decision rules for prediction.

Logistic Regression: Logistic regression is a supervised learning algorithm, use for the regression problem. It used to predict the dependent variable from a data of the collection of dependent variables. It used to predict the output of dependent categorical variable from an independent variable.

\subsection{Deep Learning Classification Models}
Deep learning algorithms are use for the classification and regression problems. Deep learning algorithms do not need the feature extraction methods, it extracts the feature from the data by itself and them build the model based on the features vector. For the deep learning models, it requires to have more data to build an efficient model.

Convolutional Neural Network: Convolutional neural network is a supervised deep learning model, it uses for both the classification and regression problems \cite{maharjan2014simple}. It has three layers, the input layer, hidden layer, and the output layer. The hidden layer performs the convolution to the input data and predict the output on the output layer.

Long Short-Term Memory: LSTM is an Artificial and deep learning algorithm use for the classification and regression problems. It is a supervised learning model; it uses the labeled data for the classification \cite{koloski2020know}--\cite{alroobaea2020deep}. It is feedback connected model. It doesn’t work on single data point but on the entire data sequence. It provides the short-term memory to the recurrent neural networks.

\subsection{Evaluation Measures}
After applying the different machine learning and deep learning models, the evaluation measures are use to evaluate the performance of the models\cite{busger2016gronup}, there are different evaluation measures used for the evaluation of the models. It includes: Confusion Matrix, Accuracy, Precision, Recall, F1-Score.

\begin{table}[htbp]
\centering
\caption{Confusion Matrix}
\begin{tabular}{@{}lcc@{}}
\toprule
 & Predicted Positive & Predicted Negative \\
\midrule
Actual Positive & TP & FN \\
Actual Negative & FP & TN \\
\bottomrule
\end{tabular}
\end{table}

Accuracy = (TP + TN) / (TP + TN + FP + FN)

Precision = TP / (TP + FP)

Recall = TP / (TP + FN)

F1-Score = (Precision * Recall) / (Precision + Recall)

\section{Results and Analysis}
In this section, the results of the proposed machine learning and deep learning models are discussed. The performance of these models is evaluated using the F1 score, precision, recall and accuracy score. The results of different models are discussed, which are applied on followers’ data to predict the demographics of celebrities.

There are four demographics that are probed for the classification of celebrity profiling on the twitter data for this research. The demographics are age, occupation,gender, and fame. For profession there are five models applied KNN have accuracy of 0.55, Logistic Regression has an accuracy of 0.60, Decision tree has accuracy of 0.40. Two deep learning modules were applied on the data, from which CNN clock the accuracy of 0.45 and LSTM had 0.50. Average of all the algorithm that is mean of accuracy is 0.525 for the profession prediction.

\begin{table}[htbp]
\centering
\caption{F1-Scores/Accuracy of machine learning models on profession}
\begin{tabular}{@{}lccc@{}}
\toprule
Model & F1-Score & Accuracy & cRank \\
\midrule
KNN & 0.55 & 0.55 & 0.588 \\
Logistic Regression & 0.60 & 0.60 & 0.601 \\
Decision Tree & 0.40 & 0.40 & 0.515 \\
Random Forest & 0.50 & 0.50 & 0.548 \\
SVM & 0.45 & 0.45 & 0.428 \\
\bottomrule
\end{tabular}
\end{table}

\begin{table}[htbp]
\centering
\caption{F1-Scores/Accuracy of deep learning models on profession}
\begin{tabular}{@{}lccc@{}}
\toprule
Model & F1-Score & Accuracy & cRank \\
\midrule
CNN & 0.45 & 0.45 & 0.515 \\
LSTM & 0.50 & 0.50 & 0.515 \\
\bottomrule
\end{tabular}
\end{table}

The second demographic of the celebrity is age prediction, which is a multiclass problem and there was total six algorithms were applied on the data for the prediction of the age. KNN has the accuracy of 0.50 for the age prediction. Logistic regression has the accuracy of 0.45 for the age prediction, Decision tree has 0.40, Random Forest 0.40 and the deep learning models CNN has 0.40 and LSTM had 0.30. The cumulative average of the age prediction accuracy is 0.475.

\begin{table}[htbp]
\centering
\caption{F1-Scores/Accuracy of machine learning models on Age}
\begin{tabular}{@{}lccc@{}}
\toprule
Model & F1-Score & Accuracy & cRank \\
\midrule
KNN & 0.50 & 0.50 & 0.564 \\
Logistic Regression & 0.45 & 0.45 & 0.539 \\
Decision Tree & 0.40 & 0.40 & 0.455 \\
Random Forest & 0.40 & 0.40 & 0.563 \\
SVM & 0.40 & 0.40 & 0.547 \\
\bottomrule
\end{tabular}
\end{table}

\begin{table}[htbp]
\centering
\caption{F1-Scores/Accuracy of deep learning models on Age}
\begin{tabular}{@{}lccc@{}}
\toprule
Model & F1-Score & Accuracy & cRank \\
\midrule
CNN & 0.40 & 0.40 & 0.455 \\
LSTM & 0.30 & 0.30 & 0.455 \\
\bottomrule
\end{tabular}
\end{table}

The third demographic which is probed is gender identification. It is a binary class problem. There is total 6 models applied for the prediction of the gender on the twitter data of the celebrities’ followers. The accuracy of the KNN is 0.55, Logistic regression has 0.60, Decision tree has 0.30, Random Forest 0.70 and the deep learning algorithms, CNN clocked 0.65 and LSTM has 0.65. The average accuracy score of the algorithms is 0.516.

\begin{table}[htbp]
\centering
\caption{F1-Scores/Accuracy of machine learning models on Gender}
\begin{tabular}{@{}lccc@{}}
\toprule
Model & F1-Score & Accuracy & cRank \\
\midrule
KNN & 0.55 & 0.55 & 0.578 \\
Logistic Regression & 0.60 & 0.60 & 0.634 \\
Decision Tree & 0.30 & 0.30 & 0.790 \\
Random Forest & 0.70 & 0.70 & 0.522 \\
SVM & 0.65 & 0.65 & 0.606 \\
\bottomrule
\end{tabular}
\end{table}

\begin{table}[htbp]
\centering
\caption{F1-Scores/Accuracy of deep learning models on Gender}
\begin{tabular}{@{}lccc@{}}
\toprule
Model & F1-Score & Accuracy & cRank \\
\midrule
CNN & 0.65 & 0.65 & 0.790 \\
LSTM & 0.65 & 0.65 & 0.790 \\
\bottomrule
\end{tabular}
\end{table}

The last demographic of the celebrity profiling is fame prediction, which is a multiclass problem. There are six models applied on the data. The classical machine learning models includes KNN has 0.30, Logistic Regression has 0.30, Decision tree has 0.25, Random forest 0.40, SVM 0.55 and deep learning algorithms CNN has 0.35, LSTM 0.45 and the average accuracy of the six algorithms is 0.358.

\begin{table}[htbp]
\centering
\caption{F1-Scores/Accuracy of deep learning models on Fame}
\begin{tabular}{@{}lccc@{}}
\toprule
Model & F1-Score & Accuracy & cRank \\
\midrule
CNN & 0.35 & 0.35 & 0.439 \\
LSTM & 0.45 & 0.45 & 0.439 \\
\bottomrule
\end{tabular}
\end{table}

There are four demographics that were probed for the classification of celebrity profiling on the twitter data for this research. The demographics are age, occupation, gender, and fame. The highest cRank among the demographics were clocked for the gender prediction which is 0.634 for the machine learning algorithms and 0.79 for the deep learning algorithms. The cRank for the gender prediction was highest because it was a binary classification problem. The lowest cRank was observed for the fame demographic which is 0.388. The fame class was highly unbalanced that’s why it didn’t show the promising results.

\section{Conclusion and Future Work}
Currently people used to communicate, interact, and build relationships through social media. Celebrities are prolific authors and most of their personal information is public knowledge. There are some digital celebrities who exist only on social media, e.g., Twitter. Twitter is a social networking service which provides general populace as well as celebrities to interact with their fans.

The demographics of celebrities could be predicted by the text of their followers as both shares same interest. However, most of the work on celebrity profiling has been performed on English and other similar languages except Urdu. On the contrary, majority of the sub-continent celebrities and their fans tweets in Urdu. To fulfill this gap, in this research work we were used Urdu tweets (short text) of 10 followers of a celebrity to build the first celebrity profiling based on followers’ tweets corpus.

There has been some issue which is faced during the research work, firstly the collection of data was a hard task for the celebrities who only tweets in Urdu and to find their followers who do the same. Secondly, for the selection of the algorithms. It was a hectic task to find out the best algorithms and the best features extraction techniques for the celebrity profiling. The feature extraction techniques include the TD-IDF and the length of the tweets,

The corpus is be preprocessed, and Machine Learning (Logistic Regression, Support Vector Machines etc.) and Deep Learning (CNN, LSTM etc.) algorithms were used to train models for the prediction task. The trained model were evaluated using state-of-the-art evaluation measures, i.e., precision, recall, and F1. The highest accuracy which is clocked is for the gender identification with 0.75 on CNN and the second best was for the KNN and it was 0.60.

\subsection{Future Work}
There are a lot of applications of the author and celebrity profiling. Author profiling playing an important role in forensics, advertisement, and the reduction of cybercrimes. This research is limited to the twitter and short text. The social media is emerging as the best tool for the assistant of the human being and in the future this research might expand to the other social media platforms include: Facebook, E-commerce Reviews profiling, LinkedIn. The research for the Urdu speaking community in these platforms will help in different fields, like to find out the fake accounts on Facebook, to probe about the fake reviews on E-commerce sites.

\bibliographystyle{plain}
\bibliography{references} % Assume a references.bib file, but since not provided, this is placeholder

\end{document}